\documentclass[conference]{IEEEtran}
\IEEEoverridecommandlockouts
\usepackage[table]{xcolor}
\usepackage{cite}
\usepackage{amsmath,amssymb,amsfonts}
\usepackage{float}
\usepackage{algorithmic}
\usepackage{graphicx}
\usepackage{lipsum}
\usepackage{capt-of}
\usepackage{cuted} 
\usepackage{comment}
\usepackage{svg}
\usepackage{textcomp}
\usepackage{booktabs}
\usepackage{ifthen}
\usepackage{multirow}
\usepackage{booktabs}
\newboolean{isAnonymous}
\setboolean{isAnonymous}{false}  

\def\BibTeX{{\rm B\kern-.05em{\sc i\kern-.025em b}\kern-.08em
    T\kern-.1667em\lower.7ex\hbox{E}\kern-.125emX}}

\newcommand\ee{\mathrm{e}}

\makeatletter
\newcommand{\newlineauthors}{%
  \end{@IEEEauthorhalign}\hfill\mbox{}\par
  \mbox{}\hfill\begin{@IEEEauthorhalign}
}
\makeatother

\begin{document}
\title{Operating Regimes of Decentralized Learning Under Mobility and Bandwidth Constraints
}

\author{
\IEEEauthorblockN{Samuele Sabella, Chiara Boldrini\IEEEauthorrefmark{2}, Lorenzo Valerio\IEEEauthorrefmark{2}, Marco Conti, and Andrea Passarella}
\IEEEauthorblockA{Institute for Informatics and Telematics, National Research Council, Italy\\
Email: \small \{samuele.sabella, chiara.boldrini, lorenzo.valerio, marco.conti, andrea.passarella\}@iit.cnr.it
}
%
}


\maketitle

\begin{abstract}
Decentralized learning is a promising paradigm for collaborative training in mobile and pervasive systems, as it avoids a central coordinator and does not require sharing raw data. Yet, most analyses rely on idealized communication assumptions that break down in wireless settings, where connectivity is intermittent, topology changes due to mobility, and bandwidth is limited. We study decentralized averaging under client asynchrony, time-varying contact graphs, and technology-dependent throughput constraints. We implement a fully decentralized protocol that overlaps synchronization with local training and supports \emph{partial} tensor-level transfers when contacts end early. Using Random Waypoint mobility and multiple wireless technologies (Bluetooth LE, LTE, and Wi-Fi), we quantify how network dynamics and link capacity impact convergence.
We identify three operating regimes: (i) inter-contact time largely dictates convergence via mixing, (ii) partial updates are often well tolerated when contacts are frequent, and (iii) very dense contact patterns can trigger contention, reducing effective throughput. These findings provide a practical lens to reason about decentralized learning deployments over realistic wireless systems, highlighting when improving connectivity, increasing bandwidth, or mitigating contention is most impactful.
\end{abstract}

\begin{IEEEkeywords}
Decentralized learning, Mobility, Time-varying networks, Wireless communication, Bandwidth constraints
\end{IEEEkeywords}

\section{Introduction}

Decentralized learning has gained increasing attention as an effective approach for collaborative model training in large-scale distributed systems. By enabling nodes to exchange model updates directly with one another, without relying on a central coordinator or sharing raw data, decentralized learning naturally addresses key challenges related to scalability, robustness, and privacy \cite{kairouz2021advances}. These properties make it particularly appealing for pervasive and edge computing scenarios, where devices are geographically distributed, heterogeneous, and often operate under limited or intermittent infrastructure support.

Despite its growing relevance, much of the existing literature on decentralized learning is grounded in idealized assumptions that limit its practical applicability. In particular, most approaches assume static or slowly varying communication topologies, reliable and homogeneous links, and abstract communication models that neglect the characteristics of real wireless technologies. In contrast, many target application domains (such as mobile sensing \cite{zhang2021mdldroid}, vehicular \cite{chen2025mobility} and drone networks \cite{qu2022decentralized}, collaborative robotics \cite{zhou2023decentralized}, and smartphone-based learning) are inherently dynamic. In these environments, \textit{nodes are mobile, connectivity is intermittent, and communication opportunities arise opportunistically based on proximity, channel conditions, and network availability}.

At the same time, decentralized learning in mobile systems must operate across a \textit{diverse set of communication technologies}, ranging from short-range interfaces such as Bluetooth and WiFi to wide-area cellular networks including LTE and 5G. These technologies exhibit fundamentally different performance profiles in terms of bandwidth, latency, reliability, energy consumption, and coverage, and therefore shape both the frequency and the effectiveness of information exchange among nodes. Abstracting away these aspects may lead to overly optimistic conclusions about learning performance and convergence, while offering limited insight into the design trade-offs faced by real systems.

Mobility and communication technologies play an especially important role in decentralized learning compared to standard federated learning (FL). In classical FL, clients primarily communicate with a fixed server through infrastructure networks, and learning proceeds in explicit rounds \cite{hard2018federated}; as a result, mobility and link heterogeneity mostly affect participation (i.e., which clients successfully upload in a given round) but do not directly reshape the aggregation structure. In decentralized learning, instead, the communication topology \emph{is} the algorithmic substrate: mobility determines which peers meet, for how long, and how quickly information can mix across the network, while the underlying wireless technology determines whether model exchanges complete, remain partial, or are delayed by congestion. Therefore, realistic mobility/contact processes and technology-specific throughput constraints can fundamentally change the effective averaging dynamics and, ultimately, the convergence behavior.

In this paper, we study decentralized learning in the presence of node mobility and realistic communication technologies. We focus on settings in which the communication graph evolves over time as a result of node movement and technology-dependent connectivity constraints, directly influencing the learning dynamics. Our goal is to better understand how mobility and heterogeneous wireless links impact the behavior, robustness, and efficiency of decentralized learning, and to narrow the gap between theoretical algorithmic models and practical deployments in mobile and pervasive environments. To better frame our problem, we introduce the following research questions addressed throughout the paper:
\begin{itemize}
    \item \textbf{RQ.1}: How do client asynchrony and time-varying connectivity (contact/inter-contact patterns) interact with link throughput to determine the convergence speed and final accuracy of decentralized averaging?
    \item \textbf{RQ.2}: How robust is decentralized averaging to \emph{partial} model transfers (i.e., truncated updates), and what fraction of an update must be received to preserve most of the learning performance?
    \item \textbf{RQ.3}: To what extent does the tensor transmission order affect convergence under bandwidth and contact-time constraints?
\end{itemize}
Beyond answering these questions empirically, our goal is to identify the system-level regimes that determine when decentralized learning is diffusion-limited, bandwidth-constrained, or contention-limited.

To answer these questions, we perform extensive simulations of DecAvg over mobility-driven contact graphs, explicitly modeling asynchronous client start times and local training durations, time-varying connectivity (contact time and inter-contact time), congestion-limited wireless throughput, and tensor-wise partial (truncated) model transfers.
Specifically, we train an image classifier on CIFAR-10 \cite{krizhevsky2009learning} with $50$ clients using Random Waypoint mobility traces, and we compare multiple communication technologies (Bluetooth LE, LTE, and Wi-Fi 4) while tracking test accuracy, update transmission percentages, and network connectivity over time.

Our study reveals three distinct operating regimes for decentralized learning under mobility and bandwidth constraints:
\begin{itemize}
    \item \textbf{Connectivity-limited regime:} When inter-contact time is large, convergence is dominated by slow information diffusion, even if links allow near-complete model transfers.
    \item \textbf{Bandwidth-tolerant regime:} When contacts are frequent and models are moderately sized, decentralized averaging remains robust to truncated exchanges; partial updates preserve most performance.
    \item \textbf{Contention-limited regime:} At very high contact rates, increased concurrency can reduce effective throughput, making intermediate mobility more favorable than extremely dense mixing.
\end{itemize}
These results provide a system-level characterization of decentralized learning beyond idealized static-topology analyses.


\section{Related work}\label{sec:related_work}

\subsection{Mobility modeling}\label{sec:related_work:sub:mobility}
Mobility models are a key component in the evaluation of wireless and mobile systems, as they determine the temporal evolution of the contact graph and therefore the opportunities for communication and information spreading. A wide range of models has been proposed, spanning simple stochastic approaches (e.g., random waypoint and its variants) to more structured models that capture correlation in user motion, spatial constraints, and social effects.

The \emph{Random Waypoint} (RWP) model remains a common choice because it provides a simple and controllable baseline in which mobility is essentially uniform and memoryless. This is particularly useful in our context because the literature quantifying the impact of mobility on decentralized learning is still relatively scarce; as a consequence, it is important to start from the simplest mobility assumptions to isolate first-order effects before moving to richer, trace-driven models. For decentralized learning, an RWP baseline enables a preliminary assessment of how \emph{uniform mixing} (i.e., contacts generated by homogeneous movement rather than by social or spatial preferences) impacts the convergence of model averaging.

At the same time, RWP is known to exhibit artifacts such as non-uniform spatial node distributions and transient effects that can bias simulation outcomes, and these limitations must be kept in mind when interpreting results\cite{bettstetter2001mobility}. Closely related baseline models include \emph{Random Direction}, which mitigates the central tendency of RWP by having nodes pick a direction and travel until they reach the simulation boundary, then select a new direction~\cite{bettstetter2001mobility}. Other stochastic models introduce additional realism by correlating motion over time; for example, the Gauss--Markov model correlates a node's velocity and direction to control the degree of randomness~\cite{liang1999gaussmarkov}.

Beyond purely random models, several proposals aim to reproduce statistical regularities observed in real traces. The \emph{Self-similar Least Action Walk} (SLAW) model captures heavy-tailed flights and pause times, as well as heterogeneity in preferred locations, producing trajectories that better resemble human movement patterns\cite{lee2012slaw}. Similarly, community- and socially-driven mobility models explicitly couple mobility with social relationships and location attraction. In particular, HCMM integrates social ties with spatial attraction and has been shown to reproduce key temporal metrics such as contact and inter-contact time distributions\cite{boldrini2010hcmm}. 

\subsection{Decentralized learning under realistic constraints}\label{sec:related_work:sub:dec}
As pervasive computing moves toward the edge, decentralized learning has emerged as a fundamental approach to exploit the available computational resources of devices outside the datacenter. The aim of decentralized learning is to collaboratively learn without relying on a centralized architecture. Unlike Federated Learning \cite{mcmahan2017communication}, which requires a central coordinator (i.e. \textit{parameter server}), fully decentralized learning allows overcoming the main limitations of a star-like topology. Namely, relying on a parameter server introduces intensive network activity towards the coordinator and a single point of failure. However, even if fully decentralized learning has gained popularity in recent years, and the effects of network topology~\cite{palmieri2024impact} or those of the initial conditions of the learning models~\cite{badie2025initialisation} have been studied, the settings under which it is examined often remain idealized. Specifically, the challenges resulting from taking into account either a dynamic network topology, client asynchronicity, or network congestion are often overlooked. Crucially, the joint impact of these variables remains altogether unexamined. For instance, while recent work by authors of \cite{chen2025mobility} explored decentralized learning with leader selection in dynamic environments, our paper proposes a leaderless scenario that jointly accounts for node mobility, network congestion, and heterogeneous training speeds. Our work is more closely related to \cite{lee2021opportunistic}, where the authors study decentralized training of image classifiers under contact duration and bandwidth constraints. The main difference is that we allow for partial updates and truncated transmissions, while their method adapts the number of local training rounds to the available communication budget.

Authors of BaCombo \cite{jiang2020bacombo} highlighted the need for decentralized learning techniques that take into account the underlying channel bandwidth. Specifically, they introduce a model sharing policy that improves bandwidth usage by fetching small chunks from many clients in parallel during each client synchronization phase. However, while they focus on maximizing throughput, our work tries to stress-test the learning algorithm under strict bandwidth constraints to evaluate the robustness of existing solutions rather than improve their efficiency. Faulty channels and unreliable communication are explored in \cite{yan2024performance}, where the impact of two distinct transmission impairments is analyzed: a) the use of UDP as the transport protocol, resulting in neighbor updates being lost with a predefined probability; b) communication over an analog channel, where received updates are corrupted by additive Gaussian noise. Although the theoretical model introduced is able to account for dynamic topologies, their empirical results consider only static topologies. Moreover, while their focus is either on missing models or corrupted updates, we assume a reliable communication protocol that is able to guarantee the integrity of the transmitted data, even though the complete model may not always be successfully delivered. In \cite{li2024decentralized}, the authors assumed a packet-level reliable communication protocol for which faulty updates are discarded. This results in some of the update parameters being set to zero, which differs from our strategy, in which we use only the received data, ignoring any unreceived portions. 

Our asynchronous approach is related to the work of SWIFT \cite{bornstein2022swift}, which proposes an asynchronous decentralized training procedure. However, our implementation is simpler since our clients use decentralized averaging without any significant changes. Similarly, DRACO \cite{jeong2024draco} proposes an asynchronous decentralized learning method aimed at reducing client idle time. DRACO assumes probabilistic training and synchronization durations, whereas we consider fixed (but client-specific) training times and explicitly simulate the network layer to account for the synchronization time. Furthermore, DRACO also assumes that some of the nodes are able to broadcast their model to the entire network during a unification step to mitigate model divergence. This assumption may be difficult to satisfy in real-world settings, as physical constraints may prevent individual nodes from communicating with all the other nodes in the network. Since we did not observe significant model drift in our experiments, we did not include such a unification step in our implementation.

\section{System Model}
\label{sec:system model}

\subsection{Decentralized learning}
\label{sec:decentralized_learning}

We model the system as a \emph{time-varying} (temporal) communication graph $\mathcal{G}^{(t)}=(\mathcal{V},\mathcal{E}^{(t)})$, where nodes are learning devices and edges represent the set of active links at communication round $t$. At each round, every device performs: (i) local training for a few epochs, (ii) transmission of its current parameters to its neighbors, and (iii) aggregation of the received parameters to update its local model.

The overall dataset is the union of disjoint local datasets,
\[
\mathcal{D} \sim \mathcal{P} = \bigcup\limits_{i=1}^{N} D_i \sim P_i \quad \text{with} \quad \bigcap\limits_{i=1}^{N} D_i = \emptyset,
\]
where $D_i$ is node $i$'s local dataset and $P_i$ its local data distribution. We assume a common model architecture and (unless stated otherwise) heterogeneous model initialization, i.e., $w_i^{0} \neq w_j^{0}
\ \forall i,j \in \mathcal{V}$ with $i\neq j$.

Following~\cite{sun2022decentralized}, let $\mathbf{w}_i^{(t)}$ denote the parameters of node $i$ at communication round $t$, and let $\ell$ be the sample-wise loss. Node $i$ performs local training on $D_i$ (e.g., a few SGD epochs on $F_i(\mathbf{w})=\frac{1}{|D_i|}\sum_{(\mathbf{x},y)\in D_i}\ell(y,h(\mathbf{x};\mathbf{w}))$) to obtain an updated local model $\tilde{\mathbf{w}}_i^{(t)}$, which it then shares with its neighbors.

Let $\mathcal{N}^{(t)}(i)$ denote node $i$ and its neighbors in the temporal graph $\mathcal{G}^{(t)}$; the model update at round $t$ is:
\begin{equation} \label{eq:aggr}
    \mathbf{w}_i^{(t)} \leftarrow
    \frac{\sum_{j \in \mathcal{N}^{(t)}(i)} |D_j| \tilde{\mathbf{w}}_j^{(t-1)}}
    {\sum_{j \in \mathcal{N}^{(t)}(i)} |D_j|}.
\end{equation}
The process repeats until convergence.
The described approach aligns with and generalizes previous work~\cite{palmieri2025robustness,valerio2023coordination,sun2022decentralized,savazzi2020federated}. Eq.~\eqref{eq:aggr} is a decentralized counterpart of FedAvg\cite{mcmahan2017communication}, with aggregation performed over time-varying neighborhoods.

\subsection{Mobility}
\label{sec:model:mobility}

We model mobility through the evolution of the temporal graph $\mathcal{G}^{(t)}=(\mathcal{V},\mathcal{E}^{(t)})$. An (undirected) edge $(i,j)\in\mathcal{E}^{(t)}$ indicates that nodes $i$ and $j$ are in contact and can exchange data during round $t$; equivalently, we define an adjacency indicator $a_{ij}^{(t)}\in\{0,1\}$. This time dependence determines the aggregation neighborhood $\mathcal{N}^{(t)}(i)$ in Eq.~\eqref{eq:aggr} and therefore the mixing pattern of model information.

From the mobility standpoint, two key statistics are the \emph{contact time} (CT) and the \emph{inter-contact time} (ICT). For a given pair $(i,j)$, CT is the duration of a contact episode, i.e., the length of a maximal time interval during which $a_{ij}^{(t)}=1$. CT bounds the amount of data that can be exchanged before the link breaks and thus, together with the technology rate, determines whether a full model can be transmitted within a single encounter (e.g., the same $10$-s contact yields very different delivered bits under Bluetooth LE vs.~Wi-Fi).
Conversely, ICT is the elapsed time between the end of a contact episode and the start of the next one for the same pair, and it controls how frequently information can be refreshed between two nodes and, more generally, how fast updates can diffuse through the temporal graph when contacts are intermittent.

\subsection{Communication technologies \& contention}
\label{sec:model:link_rate}

To capture heterogeneous technologies, each active contact is annotated with a technology $\tau_{ij}^{(t)}\in\mathcal{T}$ and a technology-dependent \emph{effective link rate} $\bar{R}_{\tau}$ (in bit/s), intended to capture the application-visible throughput after MAC effects, protocol overheads, and background traffic. 
In line with our decentralized setting, we assume \emph{direct device-to-device (D2D)} communication, i.e., links are established directly between peers without relying on a central coordinator or infrastructure-based relaying. We focus on representative technologies that are widely available on commodity devices and support peer-to-peer operation: Bluetooth Low Energy (BLE), Wi-Fi Direct (modeled via Wi-Fi 4/6 rates), and cellular sidelink communication (modeled via LTE/5G). These technologies span short-range, low-power links (BLE), medium-range high-throughput local connectivity (Wi-Fi Direct), and wide-area D2D communication supported by cellular networks. As such, they cover a broad spectrum of realistic deployment scenarios for mobile and edge devices engaging in opportunistic collaboration.
Informally, $\bar{R}_{\tau}$ is upper bounded by the nominal PHY rate for technology $\tau$.

We account for channel contention via a shared-medium policy that allocates capacity based on the instantaneous source load. Let $\mathrm{load}_i^{(t)}$ denote the number of simultaneous transfers involving node $i$ at time $t$ (according to the channel model). The instantaneous bandwidth assigned to a message sent from $i$ to $j$ is
\vspace{-5pt}
\begin{equation} \label{eq:bandwidth} 
    b_{i\rightarrow j}^{(t)} = \frac{\bar{R}_{\tau_{ij}^{(t)}}}{\mathrm{load}_i^{(t)}}.
\end{equation}
While Eq.~\eqref{eq:bandwidth} is intentionally simple, it matches our abstraction level: we study how limited contact opportunities and contention reduce the rate available to exchange (possibly truncated) updates, rather than modeling a specific MAC/PHY. By absorbing protocol overheads and background usage into $\bar{R}_{\tau}$ and representing contention through $\mathrm{load}_i^{(t)}$, the model captures the key first-order effect—each node’s available transmission time/capacity must be split among simultaneous transfers—while remaining technology-agnostic and enabling controlled, repeatable experiments across heterogeneous link settings.

\subsection{Handling of partial updates}
\label{sec:model:partial_updates}

Finally, mobility and contention may prevent a full model update from being delivered within a single contact. Let $S$ be the model size (in bits) and let $p_{i\rightarrow j}(t)\in[0,1]$ denote the fraction of the current update sent from node $i$ to node $j$ that has been received by time $t$. During a contact interval of length $\Delta t$ (i.e., when $a_{ij}^{(t)}=1$), the reception progress evolves as
\vspace{-5pt}
\begin{equation}\label{eq:progress}
    p_{i\rightarrow j}(t+\Delta t)=\min\left\{1,\;p_{i\rightarrow j}(t)+\frac{b_{i\rightarrow j}^{(t)}\,\Delta t}{S}\right\}.  \vspace{-2pt}
\end{equation}
If node $j$ performs aggregation at time $t$ before the transfer completes, it can still use the partially received update: we map $p_{i\rightarrow j}(t)$ to a tensor-wise prefix (based on the chosen tensor-priority policy) and aggregate only the tensors that have been fully received. In this case, any remaining in-flight portion of the update is discarded. Conversely, if the contact terminates before $j$ aggregates, the transfer state is preserved and may resume from $p_{i\rightarrow j}(t)$ when the two nodes meet again.

\subsection{Mobility-driven asynchronous model dissemination}
\label{sec:model:abc}

To extend decentralized learning to dynamic, bandwidth-constrained environments, we introduce several key modifications to the three-stage learning algorithm executed by each client (Sec.~\ref{sec:decentralized_learning}).
Clients are designed to share the parameters of their most recent model in parallel with aggregation and local training. Each client continuously transmits its latest model version while receiving updates from its current neighborhood. At the end of each local training cycle, assumed to take a non-negligible time $\Delta T_{tr}^i$ for node~$i$, the local model is aggregated with the updates received from neighbors according to Eq.~\ref{eq:aggr}. The resulting parameters $w_i^{(t)}$ are then made available for dissemination while the next training cycle starts in the background. While both model sharing and local training require non-negligible time, aggregation is assumed to be instantaneous.

To promote information flow across the network, clients do not necessarily wait for the end of a local training cycle to communicate. Instead, whenever a new link is established, the two clients immediately begin exchanging their latest parameters—i.e., those produced at the end of the most recently completed training cycle.


\section{Experimental settings}
\label{sec:experimental_settings}

We now describe the experimental setup used to evaluate decentralized learning under realistic mobility- and bandwidth-constrained communication.

\textit{Task, dataset, and model.}
Our main experimental setup trains an image classifier on the CIFAR-10 dataset using a network of $50$ clients. Compared to simpler benchmarks such as MNIST, CIFAR-10 is substantially more challenging due to its higher visual complexity and variability. While MNIST can be learned with very small neural networks (a reference parameter count is reported in Table~\ref{table:mnist_cifar}), CIFAR-10 typically requires larger models to achieve competitive accuracy. 

For CIFAR-10, we employ a lightweight convolutional neural network, with residual blocks and the GeLu activation function \cite{hendrycks2016gaussian}. Each block implements the transformation $y=x + \gamma \cdot F(x)$, where $\gamma$ is a learnable scaling parameter initialized with a small value and each transformation function $F$ is defined as $Conv. \rightarrow Layer\ Norm. \rightarrow GeLu \rightarrow Dropout \rightarrow Conv.$
The classification head is a single linear layer. Optimization is performed using stochastic gradient descent with a learning rate of $0.05$ and momentum of $0.9$, and we employ a batch size of $128$ samples.

\begin{table}[t]
\centering
\caption{Comparison of MNIST and CIFAR-10 datasets and corresponding model sizes}\label{table:mnist_cifar}
\begin{tabular}{lrr}
    \toprule
\textbf{Attribute} & \textbf{MNIST} & \textbf{CIFAR-10} \\
\midrule
Samples & $60,000$ & $50,000$ \\
Image size & $28\times 28$ & $32\times 32$ \\
Parameters & $21,840$ & $153,514$ \\
Expected Size (MBit) & $0.69$ & $4.91$ \\
Test accuracy (centralized) & 0.988 & 0.805 \\
\bottomrule 
\end{tabular} \vspace{-15pt}
\end{table}

\textit{Communication technologies and bandwidth model.}
To make the communication setting explicit, we report the physical-layer theoretical peak data rates (PHY) for each wireless technology in Table~\ref{table:PHYs}. We model constant background traffic by halving the nominal peak PHY rate, so that only $50\%$ of the theoretical capacity is available for model transfers. We apply this $50\%$ factor to obtain the effective per-link rate $\bar{R}_{\tau}$, which is then further divided by the instantaneous source load in Eq.~\ref{eq:bandwidth}.
Since very high-throughput technologies (e.g., Wi-Fi~8) are unlikely to be limiting in our scenarios, we focus on technologies that can become bottlenecks in practice, namely Bluetooth Low Energy, LTE, and Wi-Fi~4. Under these assumptions, the MNIST model can be exchanged with negligible overhead across all technologies, whereas the larger CIFAR-10 model may dominate the time budget, motivating an explicit treatment of bandwidth constraints. Table~\ref{tab:trasfertime_contention} reports the expected model transfer times under different technologies and source load levels, highlighting how contention can quickly dominate the communication budget for moderate-size models such as CIFAR-10.

\begin{table}[t]
\centering
\caption{Recent wireless communication technologies and their maximum theoretical data rates \cite{tplink_wifi8}\cite{etsi_4g}\cite{bluetooth_spec}\cite{etsi_5g}}
\label{table:PHYs}
\begin{tabular}{lr}
\toprule
Technology & Max PHY rate (Mbps) \\
\midrule
5G & $2\ee 4$ \\
Wi-Fi 6 & $9.6\ee 3$ \\
Wi-Fi 4 & $6\ee 2$ \\
LTE & $50$ (uplink) \\
Bluetooth 5.4 LE 1M & 1 \\
\bottomrule
\end{tabular}
\vspace{-10pt}
\end{table}

\begin{table}[t]
\centering
\scriptsize
\setlength{\tabcolsep}{3pt}
\renewcommand{\arraystretch}{1.05}
\caption{Expected model transfer time (s) for different wireless technologies and source loads.}
\label{tab:trasfertime_contention}
\begin{tabular}{llrrrrr}
\toprule
\multirow{2}{*}{Tech. (halved)} & \multirow{2}{*}{Data} & \multicolumn{5}{c}{$load_i^{(t)}$} \\
\cmidrule(lr){3-7}
 &  & 1 & 10 & 30 & 90 & 200 \\
\midrule
BT 5.4 LE 1M & MNIST & 1.398 & 13.978 & 41.933 & 125.798 & 279.552 \\
 & CIFAR-10 & 9.825 & 98.249 & 294.747 & 884.241 & 1964.979 \\
\midrule
LTE & MNIST & 0.028 & 0.280 & 0.839 & 2.516 & 5.591 \\
 & CIFAR-10 & 0.196 & 1.965 & 5.895 & 17.685 & 39.300 \\
\midrule
Wi-Fi 4 & MNIST & 0.002 & 0.023 & 0.070 & 0.210 & 0.466 \\
 & CIFAR-10 & 0.016 & 0.164 & 0.491 & 1.474 & 3.275 \\
\midrule
Wi-Fi 6 & MNIST & 0.000 & 0.001 & 0.004 & 0.013 & 0.029 \\
 & CIFAR-10 & 0.001 & 0.010 & 0.031 & 0.092 & 0.205 \\
\midrule
5G & MNIST & 0.000 & 0.001 & 0.002 & 0.006 & 0.014 \\
 & CIFAR-10 & 0.000 & 0.005 & 0.015 & 0.044 & 0.098 \\
\bottomrule
\end{tabular}
\vspace{-15pt}
\end{table}

\textit{Mobility and contact patterns.}
To model mobility-driven network dynamics, we generate contact patterns using the Random Waypoint model. Although RWP is idealized (see Sec.~\ref{sec:related_work:sub:mobility}), it provides a simple and widely adopted baseline that allows controlled manipulation of contact and inter-contact statistics. We adopt it to isolate the effects of time-varying connectivity and bandwidth constraints without introducing spatial or social biases. While richer mobility models or real traces may shift quantitative thresholds, the qualitative distinction between connectivity-limited, bandwidth-tolerant, and contention-limited regimes arises from the interaction between mixing frequency and link capacity and is therefore expected to generalize.
We tune the RWP parameters to obtain target average contact times (CT) abd inter-contact times (ICT), considering CT $\sim4$~s and ICT values between $\sim25$~s and $\sim20{,}000$~s. The choice CT $\sim4$~s represents a short proximity window typical of opportunistic pedestrian encounters and serves as a stress condition: some technologies can complete a full model transfer within a contact, while others cannot, making partial exchanges frequent and exposing bandwidth limitations.
To achieve these values, we fix the number of nodes ($50$), mobility speed ($[1,1.86]$\,m/s), connection threshold ($\sim10$\,m), and zero pause time, and vary only the simulation area to modulate node density and thus control ICT. This setting captures intermittent proximity-based interactions that directly shape both per-contact data exchange and information diffusion speed.

\textit{Synchronization policy.}
Because limited bandwidth and short contacts may prevent the complete transmission of a model update, we synchronize parameters at the tensor level (i.e., we transmit the model as a sequence of layer-wise parameter tensors) and use a fixed transmission order. This enables \emph{priority-based} sharing policies (e.g., transmitting the feature extractor first or, conversely, the classifier head first), which determine which parts of the model are more likely to be received when a transfer is truncated. Unless otherwise stated, our default policy prioritizes the feature-extractor layers.

\textit{Training and evaluation protocol.}
Clients start asynchronously at random times uniformly drawn between $0$~s and $15$~s. Each local training cycle lasts between $5$~s and $8$~s, depending on the client. Model evaluation is performed every $40$~s, corresponding to approximately six local training cycles. The training data are distributed in a non-IID fashion across clients according to a Zipf distribution with parameter $\alpha=3.5$.



Each client trains locally on a non-overlapping partition of the dataset using stochastic gradient descent and a shared model architecture, with heterogeneous initializations. Training runs for $15$ epochs with batch size $128$, learning rate $0.05$, and momentum $0.9$. To prevent local performance degradation from propagating through aggregation, we adopt early stopping: if local validation performance decreases, the client retains the previous model. Early stopping is applied locally using a small validation split from each client's data and only reverts to the previous local checkpoint; it does not introduce additional communication or coordination.

For each experiment, we track: (i) the global test accuracy (assuming a publicly available test set), (ii) the fraction of each model update successfully received by clients, and (iii) the number of active links in the network over time. 

\section{Results}
\label{sec:results}

We evaluate decentralized averaging under (i) different wireless technologies (affecting throughput), (ii) different inter-contact times (ICT, affecting connectivity), and (iii) different tensor-priority policies for partial transmissions. We report global test accuracy together with transmission dynamics and highlight the regimes where topology vs.\ bandwidth is the main limiting factor.

\subsubsection{Baselines}
\label{sec:baselines}

\begin{figure}[t]
    \centering
    \includegraphics[width=0.8\linewidth]{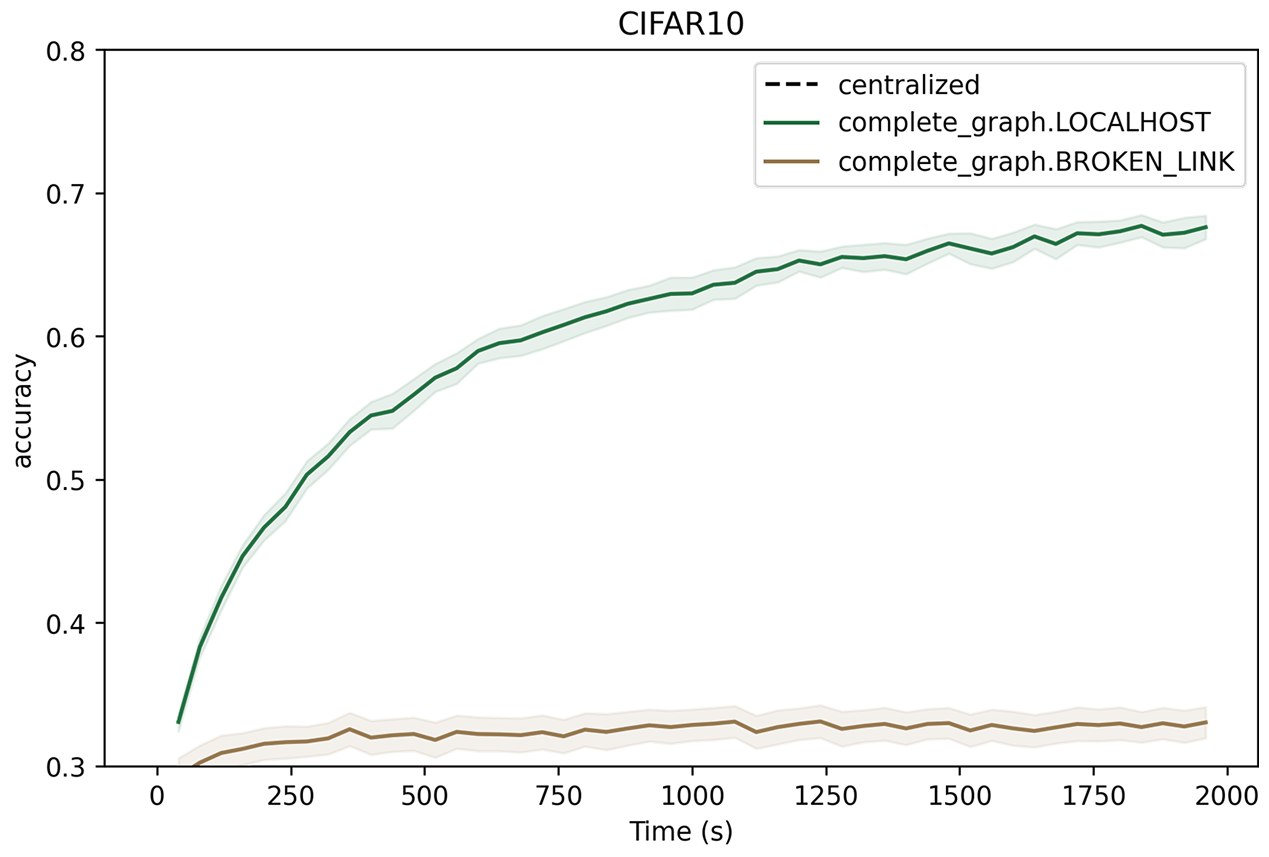} \vspace{-10pt}
    \caption{Complete-graph benchmark (continuous all-to-all connectivity; mobility has no effect) with either infinite throughput (LOCALHOST) or no communication (BROKEN\_LINK).}
    \label{fig:benchmark} \vspace{-15pt}
\end{figure}

We start with an ideal benchmarking setting: a complete-graph topology where every pair of clients is continuously connected, so mobility has no effect. Links have either infinite throughput (LOCALHOST) or zero throughput (BROKEN\_LINK); here, LOCALHOST acts as an upper bound in which communication is effectively instantaneous and never fails (as if clients were connected on the same machine, hence the name), while BROKEN\_LINK provides a lower bound with no communication. We compare both cases against a fully centralized baseline, i.e., a standard server-coordinated training setup in which a single entity holds all data and trains on it without peer-to-peer connectivity constraints, providing a reference upper bound on attainable performance. As shown in Fig.~\ref{fig:benchmark}, our decentralized runs do not fully match centralized performance within the simulated time horizon, primarily because decentralized averaging requires longer mixing time to propagate information across nodes. Extending the simulation time narrows this gap but increases computational cost substantially. We therefore stop simulations between $2{,}000$ and $3{,}000$ virtual seconds, which provides a reasonable trade-off between execution time and attained accuracy. When clients cannot exchange updates (BROKEN\_LINK), the network quickly plateaus at about $30\%$ accuracy. When updates are delivered successfully (LOCALHOST), the network learns and reaches up to $70\%$ accuracy.

\subsubsection{Technology-Dependent Bandwidth Under Fast vs.\ Slow Mixing Dynamics}
\label{sec:results_tech_ictextremes}

\begin{figure*}
    \centering
    \includegraphics[width=0.8\linewidth]{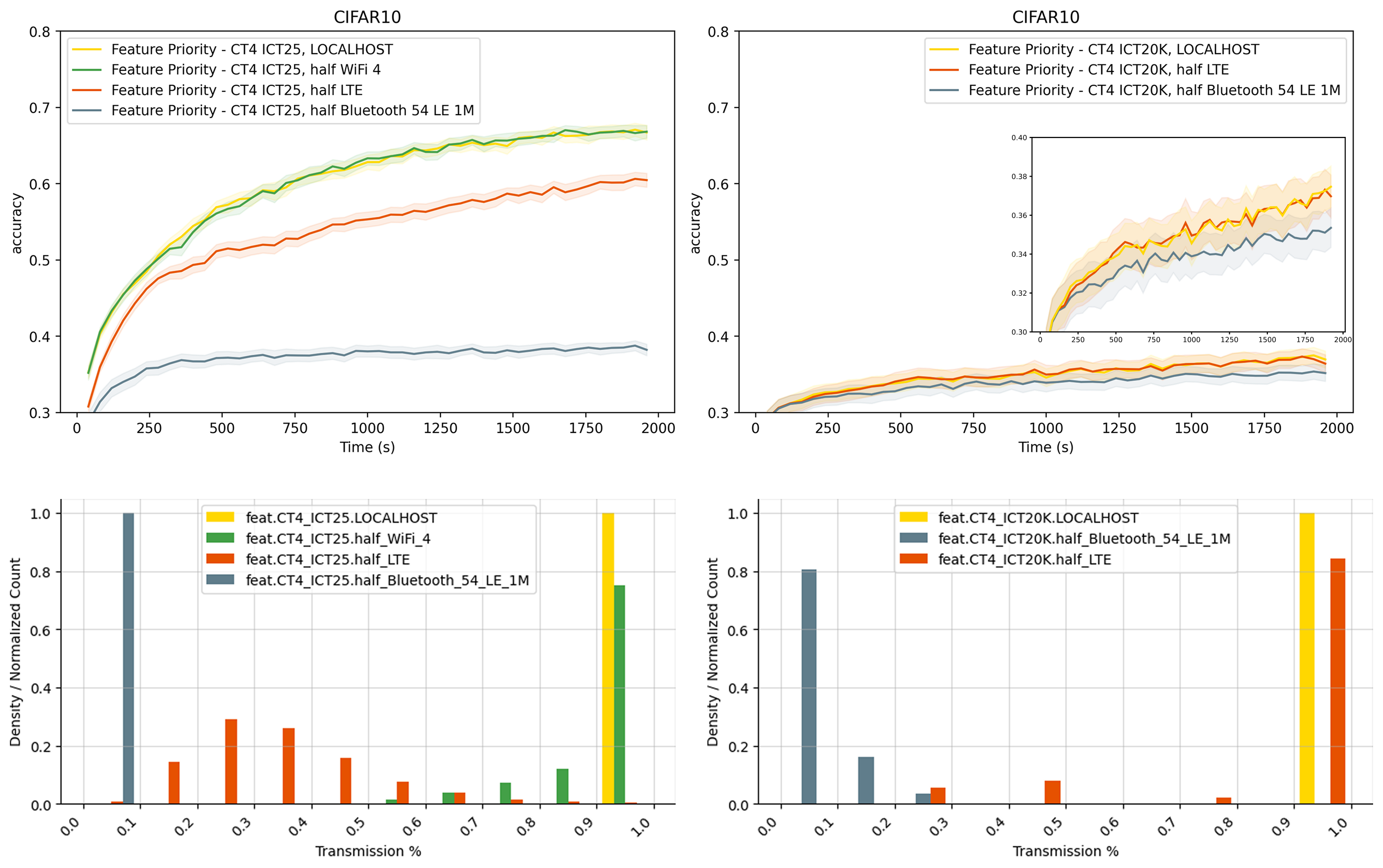}\vspace{-10pt}
    \caption{Accuracy and transmission outcomes across technologies under two extreme inter-contact-time (ICT) regimes: ICT $=25$~s (frequent contacts) and ICT $=20{,}000$~s (rare contacts). LOCALHOST is retained as an ideal reference. For readability, we omit configurations that achieve $100\%$ transmission success throughout the run (Wi-Fi 6 and 5G for ICT $=25$~s; Wi-Fi 4, Wi-Fi 6, and 5G for ICT $=20{,}000$~s). Top-left: CIFAR-10 test accuracy vs.\ time for ICT $=25$~s. Top-right: CIFAR-10 test accuracy vs.\ time for ICT $=20{,}000$~s (inset highlights the high-accuracy configurations). Bottom-left: distribution of successful transmission fractions for ICT $=25$~s. Bottom-right: distribution for ICT $=20{,}000$~s. }
    \label{fig:ICT32 vs ICT20K} \vspace{-20pt}
\end{figure*}

After establishing a baseline, we study how \emph{partial} model transmissions affect convergence. We run the same learning protocol while varying the communication technology (hence throughput) and the mobility regime (hence contact opportunities). Note that we focus on a contact time (CT) of about $4$~s, which acts as a useful stressor for the considered technologies: for some links, the contact duration is, on average, insufficient to complete a full model transfer, resulting in transmission success rates below $100\%$. Figure~\ref{fig:ICT32 vs ICT20K} compares two extreme \textit{average} inter-contact-time (ICT) settings: ICT $=25$~s, which yields a highly connected and dynamic topology, and ICT $=20{,}000$~s, which yields an almost disconnected graph (at most one contact per pair over the whole experiment).

When updates are transmitted almost completely (e.g., Wi-Fi 4 with ICT $=25$~s), accuracy matches the unlimited-bandwidth reference, indicating that full-model transmissions are not strictly necessary in this regime. This suggests that, in highly connected settings, a bandwidth-aware policy that deliberately transmits only a subset of the model parameters (rather than the full update at every contact) could preserve convergence performance while reducing communication overhead. Even when only a small fraction of the model is exchanged, the degradation is limited (e.g., LTE with ICT $=25$~s). This effect is most visible with Bluetooth: despite its very low rate, the system still improves slowly rather than plateauing as in the no-communication BROKEN\_LINK case (Fig.~\ref{fig:benchmark}).

In contrast, when the time-varying contact graph is sparse (as in the ICT $=20{,}000$~s case), contacts are so rare that most pairs of nodes do not meet (or meet only once) during the experiment. As a result, information diffuses slowly across the network, and convergence becomes severely limited even if the few established links allow nearly complete transmissions when they occur (e.g., LOCALHOST and LTE). Overall, these results indicate that under extreme ICT values, \emph{connectivity over time} dominates convergence: frequent contacts (even if some transfers are truncated) enable faster mixing than rare contacts, even when bandwidth is high.

\subsubsection{From Sparse to Dense Mixing: the Impact of Contention} 
\label{sec:results_ict}

\begin{figure*}
    \centering \vspace{-20pt}
    \includegraphics[width=0.8\linewidth]{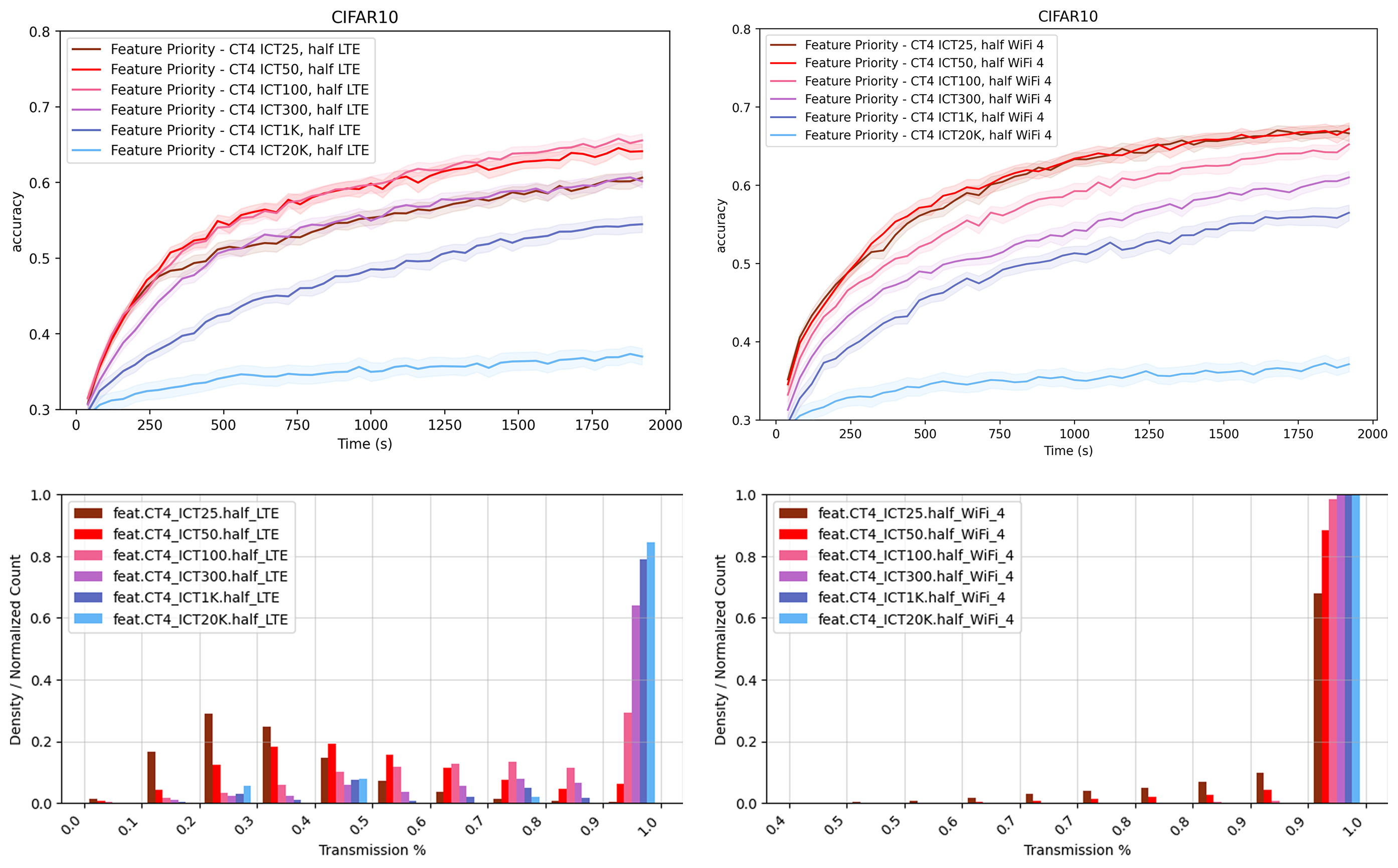}\vspace{-10pt}
    \caption{Impact of inter-contact time (ICT) on convergence for fixed wireless technologies (LTE and Wi-Fi 4), highlighting the transition from sparse mixing to dense mixing where channel contention can emerge. Top-left: CIFAR-10 test accuracy vs.\ time under LTE for different ICT values. Top-right: CIFAR-10 test accuracy vs.\ time under Wi-Fi~4. Bottom-left: distribution of successful transmission fractions under LTE. Bottom-right: distribution under Wi-Fi~4.  }
    \label{fig:ICT comparisons}
\end{figure*}

The two extreme ICT regimes considered above (very frequent vs.\ very rare contacts) capture the endpoints of the mixing spectrum, but they are not sufficient to characterize the \emph{transition} from sparse connectivity to dense connectivity, where increased contact opportunities can also trigger channel contention. To study this intermediate regime, we vary ICT while keeping the communication technology fixed. We focus on LTE and Wi-Fi 4 and omit technologies that consistently allow full-update transfers. Moreover, we exclude Bluetooth, as its limited bandwidth led to consistently poor performance in preliminary tests. Results are shown in Fig.~\ref{fig:ICT comparisons}.

Under Wi-Fi 4, throughput is generally sufficient for most updates to reach their destination, so increasing ICT (i.e., making contacts less frequent) slows convergence monotonically. Under LTE, instead, decreasing ICT too much can increase contention and reduce the effective fraction of successfully transmitted updates: ICT $=25$~s performs similarly to ICT $=300$~s, and both are outperformed by ICT $=100$~s.


\subsubsection{Tensor-Priority Policy Ablation}
\label{sec:tensor_priority}

\begin{figure}
    \centering
    \includegraphics[width=0.8\linewidth]{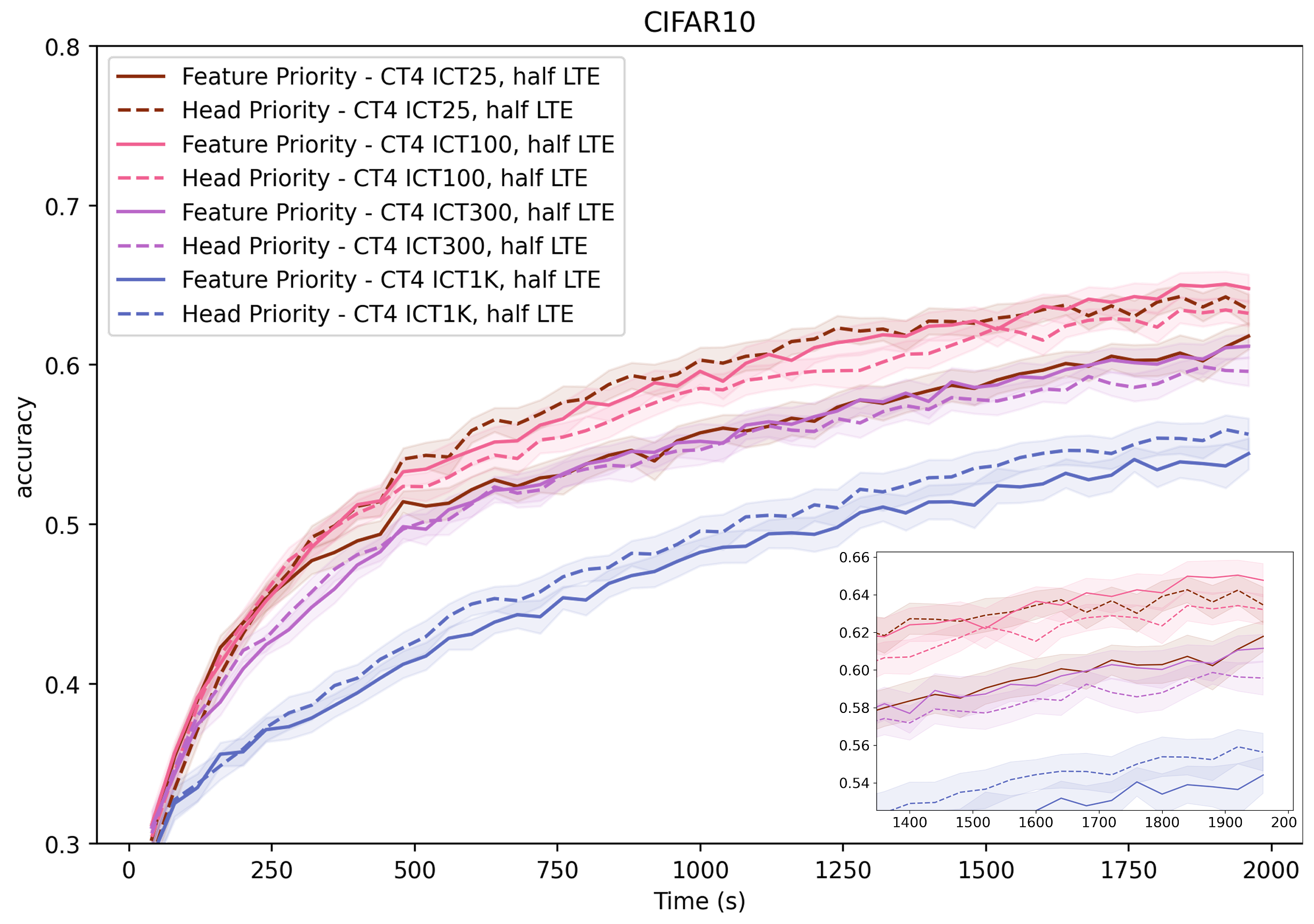} \vspace{-10pt}
    \caption{Effect of tensor transmission order under LTE: prioritizing feature-extraction tensors vs.\ classifier-head tensors.} \vspace{-10pt}
    \label{fig:policy_comparison}
\end{figure}

As a final experiment, we investigated the impact of the tensor-priority policy adopted during the update transmission.  To isolate effects due solely to network communication (rather than unequal data volumes), we ignore the amount of data held by each node when averaging models (i.e., in Eq.~\ref{eq:aggr} we do not weight updates by local dataset size). 
In Fig.~\ref{fig:policy_comparison}, we compare our default policy, which prioritizes tensors corresponding to feature extraction layers, with the opposite strategy, where layers closer to the classifier head are sent first. Under moderate communication constraints, the two policies achieve similar performance, indicating that transmission order is not a dominant factor.
When network congestion becomes severe (e.g., ICT $=25$~s), however, the choice of priority matters: sending classifier-side tensors first is beneficial, while prioritizing feature-extraction tensors can be counterproductive. A plausible explanation may be that the classifier head is smaller and aligns faster across nodes, and thus it can be synchronized more reliably within short contacts. Faster head alignment translates into earlier agreement on decision boundaries, which is particularly valuable when updates are frequently truncated.


\section{Conclusion}\label{sec:conclusion} \vspace{-3pt}
In this paper, we investigated the impact of dynamic network topology and limited channel bandwidth on decentralized averaging, focusing on scenarios where clients receive partial model updates due to mobility and communication constraints. Our simulations account for heterogeneous local training durations and asynchronous initialization. 
Using CIFAR-10 image classification, we evaluated multiple wireless technologies and mobility patterns. Results show that inter-contact time is the main factor affecting convergence speed, while partially received updates and simple tensor-priority policies have limited impact on performance.
Although our analysis focuses on moderate-scale convolutional models, the distinction between connectivity-driven diffusion and bandwidth-driven truncation is expected to generalize to larger settings. Extending the study to larger architectures and real-world traces is left for future work.

\vspace{-10pt}
\section*{Acknowledgment}
This work was partially supported by the PNRR Project ``SoBigData.it'' (IR0000013). S. Sabella, C. Boldrini, and M. Conti were partly funded by the PNRR project “FAIR” (PE00000013), while A. Passarella and L. Valerio were partially supported by the PNRR project ``RESTART'' (PE00000001).
\vspace{-10pt}


\bibliographystyle{IEEEtran}
\bibliography{references}

\end{document}